\documentclass{elsart}
\usepackage{graphicx,amssymb}

\begin{document}

\begin{frontmatter}

\title{On time and ensemble averages in quasistationary state of
  low-dimensional Hamiltonian maps}

\author{Fulvio Baldovin}
\ead{baldovin@cbpf.br}

\address{Centro Brasileiro de Pesquisas F\'{\i}sicas,
Rua Xavier Sigaud 150, Urca 22290-180 Rio De Janeiro -- RJ, Brazil 
}

\begin{abstract}
We discuss the relation between ensemble and time averages
for quasistationary states of low-dimensional symplectic
maps that present remarkable analogies with similar states
detected in many-body long-range-interacting Hamiltonian systems.
\end{abstract}


\end{frontmatter}

\section{Introduction}
A possible approach to complex phenomena, when complete
solutions are not available, is to reduce the problem into a
simpler one that still retains the important features of
interest, and to use the simplified version as a model that
gives some insight about the original issue.  Along these
lines low-dimensional maps are useful tools, widely used in
the literature, both to study specific behaviors of
nonlinear systems with few degrees of freedom, and to
understand the macroscopic properties of much larger
systems.
The macroscopic behavior that we are addressing in this
paper is the emergence (depending on initial data) of
quasistationary states (QSSs) characterized by an anomalous
almost-constant value of macroscopic variables (like the
average kinetic energy), before a crossover to 
equilibrium. 
These QSSs where for example displayed in
the dynamics of 
long-range $N$-body Hamiltonian Mean Field (HMF) models 
where it has been shown the emergence of long-standing
phases with non-Gaussian velocity distributions 
and with an anomalous value of the temperature
\cite{tsallis_00}, before the Boltzmann-Gibbs (BG)
equilibrium is attained. 
The physics associated with these QSSs is extremely rich, for
example various connections with nonextensive statistical
mechanics \cite{tsallis_00}, with
aging \cite{montemurro_01} and glassy
dynamics \cite{pluchino_01} have been pointed out.

Using a properly defined `dynamical
temperature' in coupled symplectic maps,
it is possible to exhibit QSSs similar to those observed in
the HMF model \cite{baldovin_00,baldovin_01}. 
In fact, the presence of well known mechanisms (see,
e.g., \cite{mackay_01,kaneko_01})  that are related to the 
Kolmogorov-Arnold-Moser (KAM) 
theory provides a possible theoretical background for the explanation of 
such anomalous phases.
Of course, the new interest in these results stems from the
possibility of
better understanding the dynamical mechanisms that lead to 
violations of the classical BG theory, such as the one
discussed in \cite{tsallis_00}.  
In this paper we restrict to low-dimensions, addressing  
the connection between ensemble and time averages. 
The former were already studied in \cite{baldovin_00};
the latter are obtained here following
the dynamical behavior of a single trajectory,
in a similar way as a thermometer time-integrates the interactions
with the system to provide the output temperature.

\section{QSSs using ensemble averages}
\label{section_QSS}
Symplectic maps are convenient tools for the 
study of Hamiltonian systems, since
they are the result of a Poincar\'e section in the phase space 
of a Hamiltonian system.
As pointed out in \cite{baldovin_00}, QSSs similar to those
detected in the HMF model can be reproduced in low-dimensional
symplectic maps using ensemble averages. 
In this section we report the main results.

\begin{figure}
\begin{center}
\includegraphics[width=10cm,angle=0]{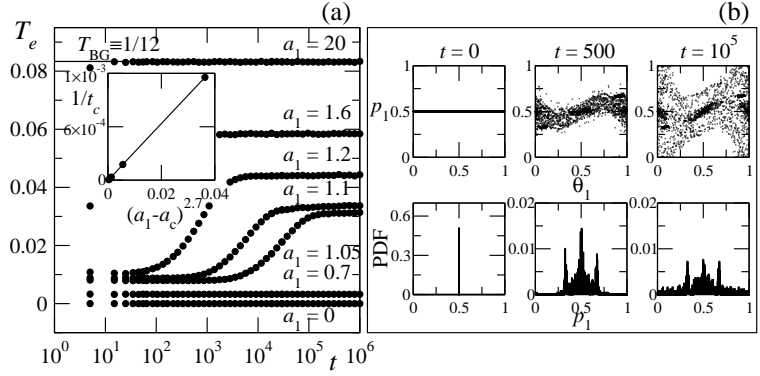}
\end{center}
\caption{\small 
Case {\bf (A)} `Ensemble' QSSs in the standard map.
(a) $T_e(t)$ for typical values of $a_1$. 
We start with `water bag' 
initial conditions ($M=2500$ points in  $0\leq\theta_1\leq 1$,
$p_1=0.5\pm 5\;10^{-4}$). 
{\bf Inset:} Inverse crossover time $t_c$ (inflection point
between the QSS and the BG regime)
vs. $1/(a_1-a_c)^{2.7}$. 
(b) Time evolution of the ensemble in (a)
for $a_1=1.1$ (first row) and 
PDF of its angular momentum (second row). 
$t=0$: initial conditions; $t=t_1=500$: the ensemble 
is mostly restricted by cantori; 
$t=t_2=10^5$: the ensemble is confined inside KAM-tori.
See \cite{baldovin_00} for further details.
}
\label{fig_std_01}
\end{figure}

In connection with the KAM theory, one of the most studied
symplectic maps 
is the {\it standard map} \cite{ott_01}, which is $2$-dimensional.
Since important qualitative changes in the topology of phase
space occur when the Hamiltonian has more than two 
degrees of freedom (e.g., Arnold diffusion), it is also
important to address this more 
general context by considering for example a $4$-dimensional
{\it symplectic} map obtained by coupling two standard maps: 
\begin{eqnarray}
\begin{array}{rclr}
p_1(t+1) &
= &
p_1(t)+\frac{a_1}{2\pi}\sin[2\pi \theta_1(t)] &
\;\;\;\;\rm{(mod\;1)},\\
p_2(t+1) &
= &
p_2(t)+\frac{a_2}{2\pi}\sin[2\pi \theta_2(t)] &
\;\;\;\;\rm{(mod\;1)},\\
\theta_1(t+1) &
= &
p_1(t+1)+\theta_1(t)+b\;p_2(t+1) &
\;\;\;\;\rm{(mod\;1)},\\
\theta_2(t+1) &
= &
p_2(t+1)+\theta_2(t)+b\;p_1(t+1) &
\;\;\;\;\rm{(mod\;1)},
\end{array}
\label{standard_standard}
\end{eqnarray}
where $a_1,a_2,b\in{\mathbb R},\;\;t=0,1,...$, $b$ 
is the coupling constant, $\theta_i$ may be regarded as an
angular variable, and $p_i$ as an angular momentum ($i=1,2$).
In the following we will take into account two cases: 
{\bf (A)} for $b=0$ we consider only coordinates $1$ , that is, 
the $2$-dimensional
standard map; 
{\bf (B)} we fix $b=2$ and $a_1=a_2\equiv\tilde a$ in order to
analyze the $4$-dimensional map (see \cite{baldovin_00} for
details).  
The standard map 
is integrable when $a_1=0$, while chaoticity rapidly increases
with $|a_1|$;
we also remind that for positive $a_1$ smaller than
$a_c=0.971635406...$ the (macroscopic) chaotic sea
is disconnected, because of the presence of invariant
trajectories, called KAM-tori, that span the whole interval
$\theta\in [0,1]$. 

With some similarity with the so-called `water bag' initial
data that produce QSSs in the HMF model \cite{tsallis_00}, 
we consider at $t=0$ a 
statistical ensemble of $M$ copies of the map 
with arbitrary $\theta_i$ and $p_i$ randomly distributed
inside small 
regions of the interval $[0,1]$.
Usually, in the connection between statistical mechanics and 
dynamics for systems with diagonal 
kinetic matrix and zero average 
momentum, the temperature is set 
proportional to the average square
momentum per particle (see, e.g., \cite{livi_01}).
As we address situations with nonzero `bulk' motion,
the analogous concept, here called 
{\it dynamical temperature},
can be defined  
as the (specific) variance of the total angular momentum: 
\begin{equation}
T_e(t)\equiv \frac{1}{d/2}\sum_{i=1}^{d/2}
\left(\langle p_i^2(t)\rangle-\langle p_i(t)\rangle^2\right),
\;\;{\rm {\bf(A)}:}\;d=2,\;\;{\rm {\bf (B)}:}\;d=4, 
\label{temperature_ensemble}
\end{equation}
where $\langle\rangle$ means ensemble average.  The
qualification {\it dynamical} is used since this definition
purely descends from dynamics and not from a thermal contact
with a thermometer.  The temperature associated with the
uniform distribution in the {\it entire} phase space can be called
{\it BG temperature} and it is given, for both cases, by
$T_{\rm{BG}}\equiv
\frac{1}{d/2}\sum_{i=1}^{d/2}\left[
\int_0^1 dp_i\;p_i^2-\left(\int_0^1dp_i\;p_i\right)^2
\right]
=1/12\simeq 0.083.
$

\begin{figure}
\begin{center}
\includegraphics[width=10cm,angle=0]{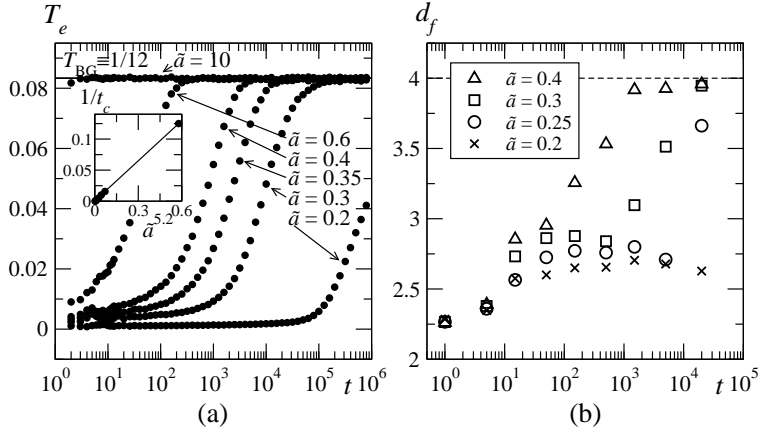}
\end{center}
\caption{\small
Case {\bf (B)} `Ensemble' QSSs in two coupled standard maps.
(a) $T_e(t)$ for $b=2$ and typical values of $\tilde a$. 
We start with `water bag' 
initial conditions ($M=1296$ points 
with 
$0\leq\theta_1,\theta_2\leq 1$, and 
$p_1,p_2=0.25\pm 5\;10^{-3}$).
{\bf Inset:} Inverse crossover time $t_c$
vs. $1/\tilde a^{5.2}$. 
(b) Time evolution of the fractal dimension 
of a single initial ensemble in the same 
setup of (a).
See \cite{baldovin_00} for further details.
}
\label{fig_std_02}
\end{figure}

For large values of $|a_1|$ (case {\bf (A)}) or $|\tilde a|$
(case {\bf (B)}) the system is typically mixing and it 
rapidly relaxes the dynamical temperature to $T_{BG}$.
Reduction of $|a_1|$ or $|\tilde a|$ causes the formation
of islands, barriers and partial barriers in phase
space. The latter are responsible for the appearance of the
QSSs, since they confine the ensemble inside a limited
volume during a certain time.
If the initial data are defined inside a singly connected
volume that does
not contain islands and if the the available portion of
phase space for the final relaxation of the ensemble is such
that its projection over the $(p_1,p_2)$-plane ($p_1$-axis
in case {\bf (A)}) is uniform, the final temperature coincides
with $T_{BG}$. We remark that this is possible only if the
dimension of the map is larger than two \cite{baldovin_00}. 
Fig. \ref{fig_std_01} and Fig. \ref{fig_std_02} display this
behavior for cases {\bf (A)} and {\bf (B)} respectively.
It is also important to notice that a nontrivial fractal
dimension characterizes the QSSs, 
while, once the cross over to $T_{BG}$ is performed, the
dimension of the ensemble coincides with that of the 
phase space (see Fig. \ref{fig_std_02} (b)).
This indicates a strong violation, during the QSSs, of the
equal-a-priori probability postulate on which BG statistical
mechanics relies.

\begin{figure}
\begin{center}
\includegraphics[width=10cm,angle=0]{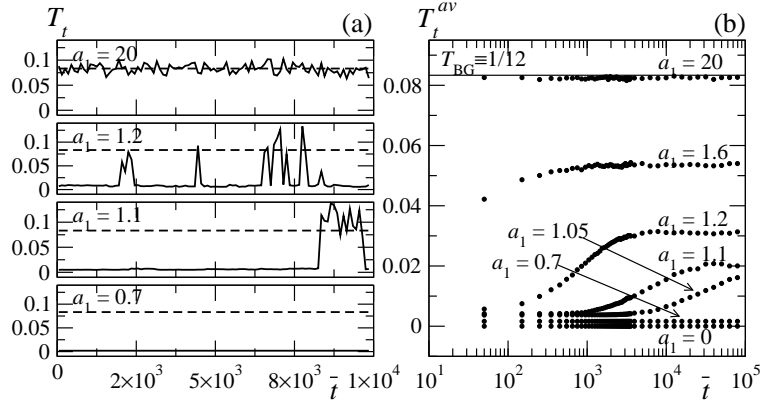}
\end{center}
\caption{\small
Case {\bf (A)} `Time' QSSs in the standard map.
(a) $T_t(\overline t)$ for typical values of $a_1$.
Evolution for a single orbit started in $x_0=(0.3,0.5)$,
with observation interval $t_o=10^2$.
Dashed lines indicate $T_{BG}$.
(b) Averages over $10^5$ realizations of the kind in (a),
with initial data inside 
$0\leq\theta_1\leq 1$: $p_1=0.5\pm 5\;10^{-4}$.
}
\label{fig_time_01}
\end{figure}

\section{QSSs using time averages}
Given a dynamical function
$f$ that takes different values $f(x_0,t)$ along a 
single  trajectory
started in the phase space point $x_0$, 
we define
its time average $\overline f(x_0,\overline t,t_o)$ at time 
$t=\overline t$ during the observation time 
$t_o\equiv t_f-t_i$ 
as
\begin{equation}
\overline f(x_0,\overline t,t_o)\equiv \frac{\sum_{t=t_i}^{t=t_f}f(x_0,t)}
{t_f-t_i},
\end{equation}
where $t_i$ and $t_f$ are respectively the initial and final 
observation time, and $\overline t\equiv t_i+t_o/2$. 
This definition mimics the action of a detector testing
the behavior of the system over a portion of a particular
trajectory.
Within this context, 
the definition of the dynamical temperature 
is reproduced as 
\begin{equation}
T_t(x_0,\overline t,t_o)\equiv \frac{1}{d/2}\sum_{i=1}^{d/2}
\left(\overline{p_i^2}-\overline p_i^2\right),
\;\;{\rm {\bf (A)}:}\;d=2,\;\;{\rm {\bf (B)}:}\;d=4. 
\label{temperature_time}
\end{equation}
For uniform distributions of momenta over the whole phase space
during the observation interval $t_o$ we obtain once again
$T_t=T_{BG}$.

\begin{figure}
\begin{center}
\includegraphics[width=10cm,angle=0]{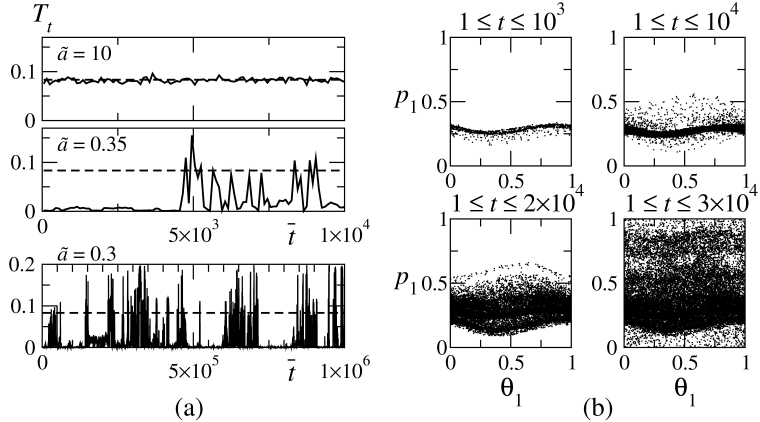}
\end{center}
\caption{\small
Case {\bf (B)}.
(a) $T_t(\overline t)$ for $b=2$ and typical values of $\tilde a$.
Evolution for a single orbit started in $x_0=(0.5,0.25,0.8,0.25)$,
with observation interval $t_o=10^2$.
Dashed lines indicate $T_{BG}$.
(b) Phase space analysis of the trajectory in (a) for 
$\tilde a=0.3$. Dots represent the projection of the
orbit on the plane $(\theta_1,p_1)$.
}
\label{fig_time_02}
\end{figure}

When the system is sufficiently chaotic, as expected, 
$T_t$ oscillates around $T_{BG}$.
On the other hand, if $|a_1|$ (case {\bf (A)}) or $|\tilde a|$
(case {\bf (B)}) are diminished, $T_t$ exhibits, instead of
the two plateaux structure, 
an {\it intermittent behavior}.
This indicates the presence of sequences of partial barriers
along the phase space trajectory, that 
alternate rapid and slow diffusion processes.
Notice that $T_t$ can jump to larger values than
$T_{BG}$ (i.e., 
the variance of the time-distribution of total momentum
can become larger than that of a uniform distribution).
Fig. \ref{fig_time_01}(a) and \ref{fig_time_02}(a) show
this behavior for cases {\bf (A)} and {\bf (B)} respectively,
for typical orbits.

\begin{figure}
\begin{center}
\includegraphics[width=10cm,angle=0]{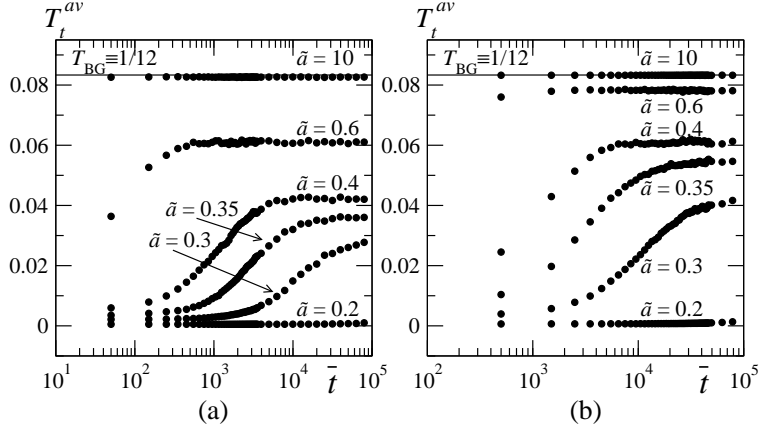}
\end{center}
\caption{\small
Case {\bf (B)} `Time' QSSs in two coupled standard maps.
$T_t^{av}(\overline t)$ obtained
averaging $10^5$ realizations of the kind in 
Fig. \ref{fig_time_02},
with initial data inside 
$0\leq\theta_1,\theta_2\leq 1$, 
$p_1,p_2=0.25\pm 5\;10^{-3}$.
(a) The observation interval is $t_o=10^2$.
(b) The observation interval is $t_o=10^3$.
}
\label{fig_time_03}
\end{figure}

Another interesting quantity is the average value of
$T_t(x_0,\overline t,t_o)$ calculated for different orbits $x_0$:
$T_t^{av}(\overline t,t_o)\equiv$\mbox{$\langle T_t\rangle_{x_0}$}.
Of course, $T_t^{av}$ coincides with $T_e$ for $t_o=1$, while
for $t_0>>1$ it represents the average result of 
different measurements and it is a typical
observable calculated in the ergodicity analysis
(see. e.g., \cite{livi_01}).
Fig. \ref{fig_time_01}(b) and \ref{fig_time_03}(a) display
that in this way we obtain a behavior qualitatively equivalent to the 
ensemble averages of Fig. \ref{fig_std_01}(a) and
Fig. \ref{fig_std_02}(a), but typically with 
{\it lower temperatures}. 
Interestingly enough, \ref{fig_time_03}(b) seems to indicate
a tendency for $T_t^{av}$ to approach $T_e$, as $t_o$
increases.

\section{Conclusions}
We have addressed a simple connection between nonlinear
dynamical systems and thermostatistics.
Using well known results in symplectic maps (see, e.g.,
\cite{mackay_01,kaneko_01}), we pointed out the presence of
QSSs with remarkable similarities with analogous ones
detected in the HMF model \cite{tsallis_00}. 
Common features include the presence of an anomalous
temperature-plateau in correspondence of particular values
of a control parameter ($a_1$ or $\tilde a$ for the maps,
the specific energy for the HMF). 
QSSs appear in both cases for particular 
classes of initial
data and after a certain
amount of time a cross over to equilibrium is observed. 
As the number $N$ of coupled
elements goes to infinity, the duration of the QSS diverges 
(see \cite{tsallis_00} for the HMF and 
\cite{baldovin_01} for the maps),
opening the possibility for this anomalous effect to be
physically relevant if the thermodynamic limit is taken
before the infinite-time limit. 
All these similarities endow with renewed interest the
analysis of dynamical mechanisms of symplectic maps. 
For example, an important result is that during the QSSs of
low-dimensional symplectic maps the phase space occupation
exhibits evidences of fractalization that disappear once
equilibrium is attained. 

The `core' of equilibrium statistical mechanics consists in 
the elimination of the time variable as can be justified
for example by the ergodic hypothesis. 
In this paper, working with low-dimensional symplectic maps, 
we have illustrated that time and ensemble averages, due to
the complexity of phase space, originate
different behaviors.
A single time average of the dynamical temperature,
a variable that was introduced in \cite{baldovin_00} in
order to enable a natural comparison with many-body
Hamiltonian systems, 
displays an intermittent behavior. 
In contrast, averaging many of these histories with a large
enough observation time we observed a two-plateaux structure
qualitatively similar to the one obtained with ensemble
averages.

{\bf Acknowledgments:} I acknowledge C. Tsallis, A. Robledo, A. Rapisarda, 
and E. Brigatti    
for useful discussions and comments, 
and M. Zamberlan for tireless encouragement.
I have benefitted from partial support by FAPERJ, CNPq and
PRONEX (Brazilian agencies).

\end{document}